\documentclass[conference]{IEEEtran}
\IEEEoverridecommandlockouts
\usepackage[T1]{fontenc}
\usepackage{cite}
\usepackage{amsmath,amssymb,amsfonts}
\usepackage{algorithmic}
\usepackage{graphicx}
\usepackage{textcomp}
\usepackage{xcolor}
\usepackage{multirow}
\usepackage{tabularx}   
\usepackage{adjustbox}  

\usepackage{tikz}
\usetikzlibrary{arrows.meta, positioning, shapes.geometric, shapes.misc, fit, calc}

\def\BibTeX{{\rm B\kern-.05em{\sc i\kern-.025em b}\kern-.08em
    T\kern-.1667em\lower.7ex\hbox{E}\kern-.125emX}}
\begin{document}

\title{Provenance Verification of AI-Generated Images via a Perceptual Hash Registry Using Merkle Patricia Trie and BK-Tree\\

}

\author{\IEEEauthorblockN{1\textsuperscript{st} Apoorv Mohit}
\IEEEauthorblockA{\textit{Market Intelligence} \\
\textit{S\&P Global}\\
Gurugram, India \\
apoorv.mohit@spglobal.com}
\and
\IEEEauthorblockN{2\textsuperscript{nd} Bhavya Aggarwal}
\IEEEauthorblockA{\textit{Market Intelligence} \\
\textit{S\&P Global}\\
Gurugram, India \\
bhavya.aggarwal@spglobal.com}
\and
\IEEEauthorblockN{3\textsuperscript{rd} Chinmay Gondhalekar}
\IEEEauthorblockA{\textit{Ratings} \\
\textit{S\&P Global}\\
New York, United States of America \\
chinmay.gondhalekar@spglobal.com}

}
\maketitle

\begin{abstract}
The rapid advancement of artificial intelligence has made the generation of synthetic images widely accessible, increasing concerns related to misinformation, digital forgery, and content authenticity on large-scale online platforms. This paper proposes a registry-based framework for verifying AI-generated images through a registry-based provenance mechanism. Each AI-generated image is assigned a digital fingerprint that preserves similarity using perceptual hashing and is registered at creation time by participating generation platforms. The Merkle Patricia Trie, provides prefix-based organization and cryptographic integrity of the registry structure, while Burkhard–Keller tree enable similarity search based on Hamming distance. Verification is performed when images are re-uploaded to digital platforms such as social media services, enabling identification of previously registered AI-generated images even after benign transformations or partial modifications. The proposed system does not aim to universally detect all synthetic images, but instead focuses on verifying the provenance of AI-generated content that has been registered at creation time. By design, this approach complements existing watermarking and learning-based detection methods, providing a platform-agnostic, tamper-evident mechanism for scalable content provenance and verification at the point of large-scale online distribution.

\end{abstract}


\section{Introduction}
The rapid rise of AI-generated content, particularly synthetic images, presents a growing challenge in distinguishing authentic visual media from artificially generated content. As text-to-image models continue to advance in realism and accessibility, the boundary between real and synthetic imagery is increasingly blurred, raising serious ethical, societal, and technical concerns related to misinformation, digital forgery, and erosion of trust in online content. Modern image synthesis systems are capable of producing highly convincing visuals that can be exploited for malicious purposes, including political propaganda, identity fraud, and manipulation of public perception. The widespread availability of these tools enables large-scale creation and mass distribution of synthetic images primarily through social media and online content-sharing platforms, where such content can rapidly propagate to millions of users. Consequently, reliable provenance verification is most critical at the point of upload and redistribution on these platforms, where misinformation spreads at scale and timely identification of AI-generated imagery is essential.

Existing approaches for identifying AI-generated images primarily rely on digital watermarking\cite{b1}, forensic analysis, or learning-based detection models \cite{b2}. Watermarking techniques embed hidden signals into images to enable later identification, while forensic methods analyze pixel-level artifacts and statistical inconsistencies introduced during image generation. More recently, AI-based detectors attempt to classify images as real or synthetic using deep learning models trained on known generators. Despite ongoing progress, these approaches face notable practical limitations. Watermarks can be degraded or removed through image transformations\cite{b3}, and learning-based detectors require frequent retraining to keep pace with evolving generative models \cite{b4}, and centralized moderation systems often operate as opaque black boxes with limited auditability. As a result, there remains a gap for a transparent, optionally decentralized, and tamper-evident mechanism that enables independent verification of AI-generated image provenance without modifying the image content or relying solely on probabilistic classification.

In this work, we address a specific and complementary problem: verifiable provenance identification of AI-generated images that have been previously registered at creation time. Unlike blind AI detectors that attempt to classify arbitrary images as real or synthetic, the proposed system verifies whether an uploaded image matches a known, registered AI-generated image or its near-duplicate variants. Consequently, this approach does not aim to universally detect all AI-generated images. Instead, it provides a deterministic and auditable mechanism for identifying re-uploads and modified instances of registered AI-generated content.

To achieve this, we propose a registry-based framework built on a Merkle Patricia Trie \cite{b5} (MPT) for tamper-resistant storage and a Burkhard–Keller \cite{b6} (BK) tree for efficient similarity search. Each AI-generated image is assigned a perceptual hash \cite{b7} (pHash) that encodes visual similarity rather than exact pixel equivalence. These hashes are registered at image creation time by participating generation platforms and stored on a public ledger. The MPT organizes hashes by prefix to enable scalable storage, while BK-trees within each prefix bucket support fast identification of perceptually similar images based on Hamming distance.

The system is designed to operate at the infrastructure level, enabling both AI image generation platforms and digital content distribution platforms to integrate provenance mechanisms into their pipelines. AI generation platforms can register content at creation time by computing the perceptual hash of each generated image and recording it on the registry. When an image is uploaded, its perceptual hash is computed and compared against the registry. If the image matches registered AI-generated content, either exactly or within a defined similarity threshold, it is flagged accordingly, along with an associated similarity score. This enables reliable identification even when images undergo benign transformations such as resizing, compression, or minor edits.

Rather than replacing existing watermarking, forensic analysis, or AI-based detection techniques, the proposed registry-based approach complements them by providing a tamper-proof and cryptographically verifiable provenance signal. Its decentralized design eliminates single points of control, ensures transparency through independent verification, and avoids intrusive modification of image content. By focusing on large-scale online redistribution rather than universal detection, the framework offers a practical and scalable solution for strengthening trust in digital imagery when it matters most at the point of widespread dissemination.

\subsection{Background}
To build a robust and scalable system for verifiable provenance identification of AI-generated images, we rely on several foundational concepts. Perceptual hashing (pHash) encodes the visual characteristics of an image into a compact 64-bit representation, enabling similarity-preserving matching that is resilient to benign transformations rather than exact pixel equivalence. Hamming distance is used as the similarity metric to quantify visual proximity between hashes. For secure, tamper-evident, and deterministic storage of registered hashes, we employ a Merkle Patricia Trie (MPT) registry. To support efficient similarity search during verification, a Burkhard–Keller (BK) tree organizes perceptual hashes within MPT buckets, enabling fast identification of registered AI-generated images and their near-duplicate variants at upload time.

\paragraph{Perceptual Hashing}
Perceptual hashing (pHash) generates a compact representation of an image by capturing its dominant visual structures rather than exact pixel values. In a typical pHash pipeline, the image is resized, converted to grayscale, and transformed into the frequency domain using a two-dimensional Discrete Cosine Transform (DCT) \cite{b8}. A subset of low-frequency DCT coefficients is then binarized relative to their median value to form a fixed-length (64-bit) hash that is robust to common image transformations such as resizing, compression, and minor brightness changes.

\paragraph{Hamming Distance}
The perceptual hash (pHash) of an image is a fixed-length binary representation that captures its visual characteristics. The similarity between two images is measured using the Hamming distance, which quantifies the number of differing bit positions between their hashes. A Hamming distance of zero indicates perceptual equivalence, while larger values correspond to increasing visual dissimilarity.

\paragraph{Merkle Patricia Trie}
A Merkle Patricia Trie (MPT) is a cryptographically authenticated key–value data structure that combines prefix-based tries with Merkle trees and is widely used in blockchain systems such as Ethereum. Keys are organized by shared prefixes, enabling efficient storage and retrieval, while cryptographic hashes at each node ensure integrity and tamper-evidence. Any modification to the stored data propagates to the root hash, allowing deterministic and verifiable validation of the entire structure. These properties make MPTs well suited for secure, scalable, and auditable storage of registered perceptual hashes.

\paragraph{Burkhard-Keller tree}
A Burkhard–Keller (BK) tree is a tree based structure designed for efficient similarity search when a well-defined distance function is available. Each node stores a value and organizes its children based on their distance from the parent, allowing queries to be restricted to a small subset of relevant branches. This pruning property significantly reduces search complexity compared to brute-force methods, making BK-trees well suited for approximate matching of perceptual image hashes using Hamming distance.

\subsection{Objectives}
The primary objective of this paper is to present a registry-based framework for verifiable provenance identification of AI-generated images registered at creation time. Using perceptual hashing (pHash), AI image generation platforms record similarity-preserving fingerprints of generated images on a public, decentralized ledger. Digital platforms can then verify whether an uploaded image matches registered AI-generated content or its near-duplicate variants by comparing perceptual hashes at upload time.

Rather than universally detecting all synthetic images, the proposed system provides a deterministic, tamper-evident, and auditable verification mechanism that complements existing watermarking and AI-based detection approaches. The goal is to enhance transparency and digital trust at the point of large-scale online distribution by offering a scalable, platform-agnostic provenance solution for AI-generated imagery.

\section{Literature Review}

The increasing realism of AI-generated media has intensified the need for reliable mechanisms to detect manipulation and establish content authenticity. Existing approaches broadly fall into three overlapping categories: AI-based detection, blockchain-enabled verification, and watermarking or provenance frameworks. While each addresses part of the problem, none offers a complete, standalone solution.

Recent research has explored hybrid AI and blockchain architectures to strengthen deepfake detection. Heidari et al. (2024) propose a blockchain-enabled federated learning framework \cite{b9} that allows multiple parties to collaboratively train deepfake detection models without sharing raw data. Their work demonstrates how blockchain can secure model updates and improve trust in decentralized learning environments. Similarly, Khalaf et al. (2024) integrate CNN-based image forgery detection with blockchain to store verification outcomes in an immutable ledger, enhancing auditability of forensic decisions \cite{b10}. These approaches primarily focus on model-centric detection, using blockchain as a secure coordination or logging layer.

Mastoi et al. (2025) shift emphasis toward content-centric provenance \cite{b11}, employing blockchain to store cryptographic hashes or watermarks of media assets. Their work highlights blockchain’s value in immutable traceability and authenticity verification, even when detection accuracy alone is insufficient. However, such systems typically assume cooperative content registration and do not address similarity-based detection of previously unseen or lightly modified media.

Beyond academic proposals, Microsoft PhotoDNA \cite{b12} represents an early and widely deployed example of perceptual hashing for content identification. PhotoDNA is highly effective at matching known images despite transformations, but it relies on centralized databases and does not inherently distinguish AI-generated content from authentic media without prior registration.

Industry-driven provenance initiatives further extend this landscape. Content Credentials (C2PA) \cite{b13} defines standardized, cryptographically signed metadata describing content origin and edits, aiming to improve transparency across platforms. Complementing metadata-based approaches, DeepMind’s SynthID \cite{b14} embeds imperceptible watermarks directly into AI-generated media, enabling origin identification even after common transformations. While both approaches strengthen disclosure and provenance, they depend on generator-side adoption and are ineffective when metadata or watermarks are stripped, absent, or never embedded.

While existing solutions significantly advance deepfake detection and content authenticity, notable gaps remain. Watermarking and provenance frameworks such as SynthID and C2PA rely on cooperative content generators and are ineffective when metadata or watermarks are absent or intentionally removed. Pure deep learning–based detectors suffer from generalization issues, dataset bias, and adversarial adaptation, limiting robustness at scale. Hash-based systems like PhotoDNA are highly effective for known content but depend on centralized registries and exact or near-exact matching. Moreover, current blockchain-based approaches primarily focus on logging detection results or model updates, lacking a unified mechanism for scalable, similarity-preserving content lookup.

\section{Methodology}
\subsection{System Architecture}

The system architecture consists of two main components:

\paragraph{Hash Generation and Storage}
AI image generators compute the perceptual hash (pHash) of each generated image and submit it to the registry.

\paragraph{Verification Process}
When a user uploads an image to a digital platform, the platform computes the pHash of the image and queries the registry to determine if it matches any registered AI-generated image.

\subsection{Image Hashing}
When an AI image generation platform produces an image, a unique perceptual hash (pHash) is generated. Unlike cryptographic hashes like SHA-256, which are highly sensitive to even the slightest pixel change, pHash is resilient to minor alterations such as resizing, brightness shifts, and compression. This tolerance makes it particularly well-suited for identifying near-duplicate AI-generated images, even if they have undergone slight modifications. Using pHash, a robust fingerprint that captures the image's perceptual identity is created.

\subsection{Storage Structure}
The resulting perceptual hash reflects the image's overall structure rather than individual pixel values. This hash is then stored in the system for similarity-based verification. Upon a new image upload, its pHash is computed and compared to the stored hashes using Hamming distance, enabling provenance of AI-generated images even when they have undergone slight modifications.
Instead of using a flat hash list, the system organizes these hashes using a Merkle Patricia Trie (MPT), where each key corresponds to a fixed-length prefix such as the first 4 hexadecimal characters or a multi segment prefix in which the prefix is calculated using 4 discontinuous digits of the 16 digit hash (e.g. concat(phash(4),phash(8),phash(12),phash(16))) of the 16 digit hex pHash. This prefix-based bucketing enables efficient insertion and retrieval by narrowing the search space to a manageable subset of visually similar images.
The Merkle Patricia Trie (MPT) holds the prefix key. The corresponding value is a reference to a BK-tree that stores all full-length pHashes sharing that prefix. Additionally, the MPT node for each prefix stores an SHA-256 hash representing the state of the associated BK-tree. This ensures cryptographic integrity and enables efficient verification. Every time a new pHash is inserted into the BK-tree, the SHA-256 hash of that tree is updated to reflect the change in structure.

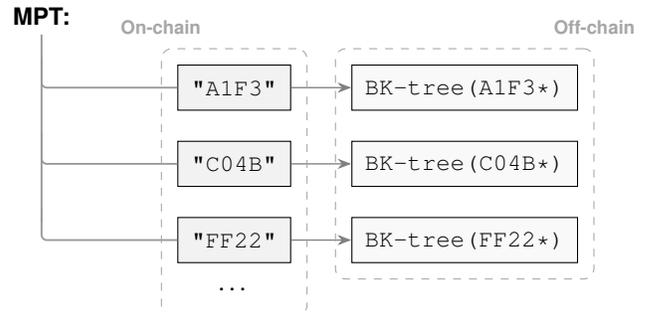
\begin{figure}[htbp]
\centering
\begin{adjustbox}{max width=\columnwidth}
\begin{tikzpicture}[
    node distance=0.4cm and 0.8cm,
    auto,
    mpt_label/.style={font=\small\bfseries\sffamily},
    key_box/.style={draw, fill=gray!10, rectangle, minimum height=0.6cm, minimum width=1.5cm, font=\small\ttfamily},
    val_box/.style={draw, fill=gray!5, rectangle, minimum height=0.6cm, minimum width=3cm, font=\small\ttfamily},
    arrow/.style={-{Stealth[scale=1.0]}, line width=0.5pt, gray},
    branch/.style={draw, gray!80, line width=0.7pt, rounded corners=2pt},
    dashedgroup/.style={draw, dashed, rounded corners=4pt, inner sep=6pt, gray!70},
    grouplabel/.style={font=\scriptsize\sffamily\bfseries, gray!70}
]

    \node (mpt) [mpt_label] {MPT:};

    \node (k1) [key_box, below right=of mpt, xshift=0.5cm] {"A1F3"};
    \node (k2) [key_box, below=of k1] {"C04B"};
    \node (k3) [key_box, below=of k2] {"FF22"};
    \node (dots) [below=of k3, yshift=0.2cm] {\dots};

    \node (v1) [val_box, right=of k1] {BK-tree(A1F3*)};
    \node (v2) [val_box, right=of k2] {BK-tree(C04B*)};
    \node (v3) [val_box, right=of k3] {BK-tree(FF22*)};

    \draw [arrow] (k1) -- (v1);
    \draw [arrow] (k2) -- (v2);
    \draw [arrow] (k3) -- (v3);

    \draw [branch] (mpt.south) -- ++(0,-2.5);
    \draw [branch] (mpt.south) ++(0,-0.3) |- (k1.west);
    \draw [branch] (mpt.south) ++(0,-1.3) |- (k2.west);
    \draw [branch] (mpt.south) ++(0,-2.3) |- (k3.west);

\end{tikzpicture}
\end{adjustbox}
\caption{Merkle Patricia Trie (MPT) structure.}
\label{fig:mpt_prefix_structure}
\end{figure}

Within each MPT bucket, a BK-tree is constructed to index the full pHashes based on Hamming distance. This structure enables fuzzy matching, allowing the system to efficiently identify near-duplicate images by exploring only relevant branches during search operations. As a result, both exact and approximate matches can be found, even at scale.

\begin{figure}[htbp]
\centering
\begin{adjustbox}{max width=\columnwidth}
\begin{tikzpicture}[
    node distance=0.25cm,
    auto,
    header_label/.style={font=\small\bfseries\ttfamily},
    hash_node/.style={draw, fill=gray!10, rectangle, minimum height=0.6cm, minimum width=3.8cm, font=\small\ttfamily},
    branch/.style={draw, black, line width=0.7pt}
]

    \node (bk) [header_label] {BK-tree (A1F3*):};

    \node (h1) [hash_node, below right=of bk, xshift=0.2cm, yshift=0.2cm] {A1F3 \textbf{0011} \dots};
    \node (h2) [hash_node, below=of h1] {A1F3 \textbf{1100} \dots};
    \node (h3) [hash_node, below=of h2] {A1F3 \textbf{0101} \dots};
    
    \coordinate (stem_x) at ([xshift=-2.5cm]h1.west);
    \coordinate (top_point) at (stem_x |- bk.south);
    \coordinate (bottom_point) at (stem_x |- h3.west);

    \draw [branch] (top_point) -- (bottom_point);

    \draw [branch] (h1.west -| stem_x) -- (h1.west);
    \draw [branch] (h2.west -| stem_x) -- (h2.west);
    \draw [branch] (h3.west -| stem_x) -- (h3.west);
    
    \node (dots) [below=of h3, xshift=-0.8cm, yshift=0.1cm] {\dots};

\end{tikzpicture}
\end{adjustbox}
\caption{BK Tree Structure}
\label{fig:bk_tree_internal_fixed}
\end{figure}
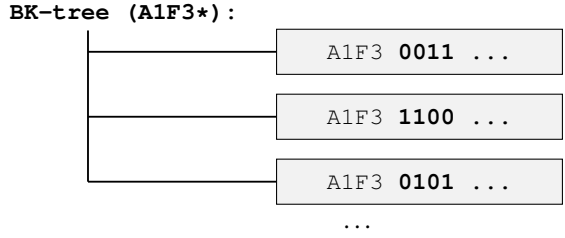

Each hash entry in the BK Tree can also include metadata such as the image creation timestamp, the platform identifier et cetera, verifying the origin of the image.

Full perceptual hashes and their associated metadata such as image creation timestamp, generating platform identifier, and optional image signatures are stored within BK-tree structures, where each node corresponds to a registered image hash. These BK-trees enable efficient similarity search based on Hamming distance and maintain all image-level contextual information.

The MPT structure is independent of any particular blockchain implementation. In deployments requiring decentralized verification, the root commitment of the MPT can optionally be anchored to a blockchain. Such integration would provide an additional trust layer while preserving the underlying registry and similarity-search architecture.

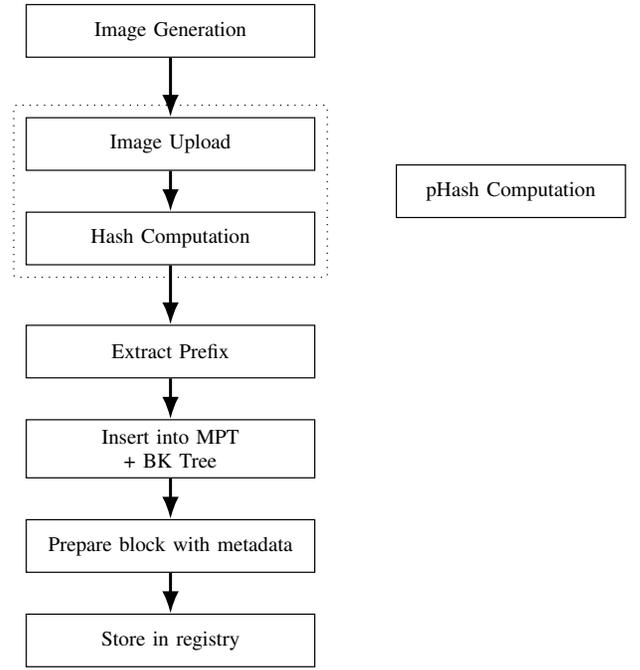
\begin{figure}[htbp]
\setlength{\fboxsep}{0pt}

\makebox[\linewidth][r]{\resizebox{0.92\linewidth}{!}{%
\begin{tikzpicture}[
  font=\scriptsize,
  node distance=4.8mm,
  box/.style={draw, minimum width=34mm, minimum height=6.2mm, align=center},
  rightbox/.style={draw, minimum width=27mm, minimum height=6.2mm, align=center},
  dottedgroup/.style={draw, dotted, rounded corners=1pt, inner sep=4pt},
  arrow/.style={-Latex, line width=0.35mm}
]

\node[box] (gen) {Image Generation};

\node[box, below=7mm of gen] (upload) {Image Upload};
\node[box, below=of upload] (hash) {Hash Computation};

\node[box, below=7mm of hash] (prefix) {Extract Prefix};

\node[box, below=of prefix] (insert) {Insert into MPT\\+ BK Tree};

\node[box, below=of insert] (prep) {Prepare block with metadata};

\node[box, below=of prep] (final) {Store in registry};

\draw[arrow] (gen) -- (upload);
\draw[arrow] (upload) -- (hash);
\draw[arrow] (hash) -- (prefix);
\draw[arrow] (prefix) -- (insert);
\draw[arrow] (insert) -- (prep);
\draw[arrow] (prep) -- (final);

\node[dottedgroup, fit=(upload) (hash)] (grp_phash) {};

\node[rightbox, right=8mm of grp_phash] (phlabel) {pHash Computation};

\end{tikzpicture}%
}}

\caption{Flowchart for Storage}
\label{fig:storage_flow_hybrid}
\end{figure}

\subsection{Search Algorithm}
When an image is uploaded to any social media platform or digital service, the platform first computes a perceptual hash (pHash) of the image. It  is then used for verification against a registry. The system begins by extracting the prefix key. This prefix maps to a specific bucket that stores full pHashes with that same prefix.
Within this bucket, a BK-tree is used to search for hashes based on Hamming distance. If a match is found with Hamming distance = 0, the image is considered an exact match and flagged as "Found in Registry". If no exact match is found within the bucket, the system expands its search to up to 137 nearby buckets, corresponding to prefixes that differ by 1 or 2 bits from the original. This accounts for minor noise or visual transformations that may slightly alter the pHash.
Each of these neighboring buckets is also searched using their respective BK-trees, again prioritizing matches with Hamming distance $= 0$. If a matching perceptual hash (pHash) is found within the search scope, the decision is based on the minimum Hamming distance between the query image and registered entries. An exact match, corresponding to a Hamming distance of zero, indicates that the uploaded image is identical at the perceptual-hash level to a previously registered AI-generated image.

For non-zero distances, the system applies an operational Hamming distance threshold $\tau$ within 137 buckets to determine whether an image should be flagged as matching registered AI-generated content. The threshold $\tau$ is selected to balance true positive and false positive rates. Distances exceeding $\tau$ are treated as non-matches.
In practical deployments, $\tau$ can be selected to satisfy platform-specific false positive rate constraints, such as limiting incorrect flagging of real images to below a predefined operational threshold.

We reach the figure of searching 137 buckets through following calculations:

{\it Case I: 1 bit flip}
Every prefix has 4 hex characters which translates to 16 bits. Therefore, mathematically,
\begin{equation}
\binom{16}{1} = 16
\end{equation}

{\it Case II: 2 bit flip}
Every prefix has 4 hex characters which translates to 16 bits. Therefore, mathematically,
\begin{equation}
\binom{16}{2} = 120
\end{equation}

Therefore, we get a total of $120 + 16 + 1 = 137$ buckets, including the original.
During a search, only 1- and 2-bit flip prefixes (up to 137 buckets) are currently considered for performance reasons. However, it is possible to extend this to include 3 and 4-bit flips, potentially increasing the false-positive rate at the cost of increased computational load. During verification, prefix search expansion is restricted to 1 and 2-bit flips, resulting in at most 137 candidate buckets per query. This design choice balances robustness to minor hash perturbations with reasonable query latency. It is important to note that the bit-flip tolerance used for prefix enumeration is independent of the Hamming distance threshold used to determine a match, allowing search scope and match strictness to be tuned separately.

\textbf{Prefix-pruning tradeoff:}
The number of prefix variations searched determines a tradeoff between robustness and search cost. Expanding the prefix search to include additional bit flips increases the number of candidate buckets examined, thereby increasing the likelihood of recovering registered images whose perceptual hashes have undergone larger prefix changes. However, the larger search scope also increases the number of candidate hashes passed to the subsequent BK-tree searches and may expose more unrelated hashes to the Hamming-distance matching criterion, increasing false-positive risk. Thus, a larger prefix-flip tolerance can improve recall at the cost of increased computational overhead and potentially higher false-positive rates. In this work, the search is restricted to 1- and 2-bit prefix flips (137 buckets including the original prefix) as a practical balance between robustness and search efficiency.

\subsection{Verification Mechanism}
If a matching perceptual hash (pHash) is found in the registry, the image is flagged based on the degree of similarity. An exact match identified by a Hamming distance of 0 is considered an exact perceptual hash match with dH = 0, and the image is conclusively flagged. This provides strong evidence that the image has previously been recorded and recognized as synthetic.
For interpretability of near-duplicate matches, we also report a normalized distance-based similarity score, defined as:

\begin{equation}
S(A, B) = \left( 1 - \frac{dH(A, B)}{n} \right) \times 100
\label{eq:similarity}
\end{equation}

Where,\\
S(A,B) = Similarity Score\\
dH(A,B) = Hamming Distance\\
n = length of the hash (64 bits)

This similarity score is a heuristic derived from normalized Hamming distance and is not intended to represent a calibrated probability of AI generation.

For example, a Hamming distance of 2 results in a 96.88\% match, while a distance of 5 yields a 92.19\% match. These images are flagged, indicating strong visual similarity to known, registered AI-generated content but allowing for slight modifications or variations.
Depending on the platform’s policy, flagged images can be accompanied by context such as similarity score, source information, or AI-generation platform details. They may also be labeled with a warning, restricted from sharing, or prioritized for manual review, helping users and moderators assess the authenticity and intent behind the content. This layered approach enhances transparency, mitigates misinformation, and promotes responsible use of synthetic media.

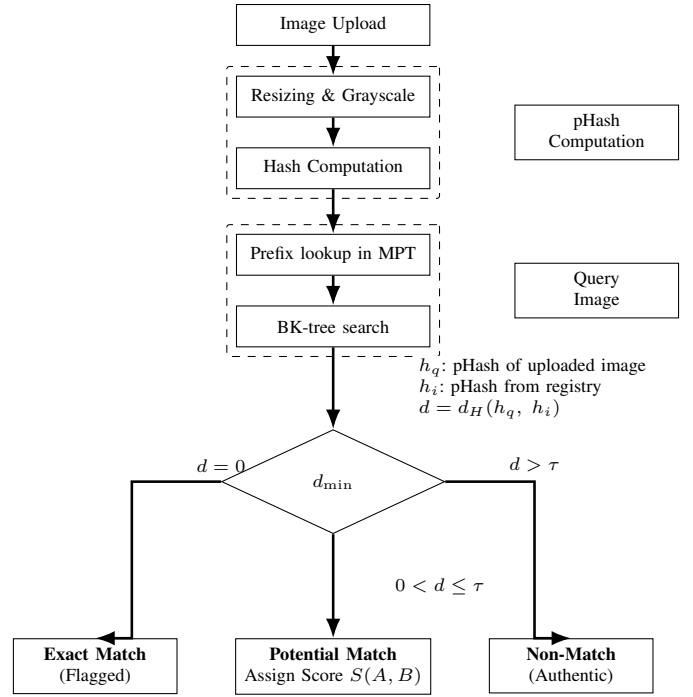
\begin{figure}[htbp]
\centering
\setlength{\fboxsep}{0pt}

\resizebox{\linewidth}{!}{%
\begin{tikzpicture}[
  font=\scriptsize,
  node distance=4mm and 7mm,
  box/.style={draw, minimum width=26mm, minimum height=5.5mm, align=center},
  sidebox/.style={draw, minimum width=22mm, minimum height=5.5mm, align=center},
  rightbox/.style={draw, minimum width=22mm, minimum height=7.0mm, align=center},
  decision/.style={draw, diamond, aspect=2.2, align=center, inner sep=1pt},
  dashedgroup/.style={draw, dashed, rounded corners=1pt, inner sep=3.5pt},
  arrow/.style={-Latex, line width=0.35mm}
]

\node[box] (upload) {Image Upload};
\node[box, below=of upload] (resize) {Resizing \& Grayscale};
\node[box, below=of resize] (hash) {Hash Computation};
\node[box, below=6mm of hash] (mpt) {Prefix lookup in MPT};
\node[box, below=of mpt] (bk) {BK-tree search};

\draw[arrow] (upload) -- (resize);
\draw[arrow] (resize) -- (hash);
\draw[arrow] (hash) -- (mpt);
\draw[arrow] (mpt) -- (bk);

\node[dashedgroup, fit=(resize) (hash)] (grp_phash) {};
\node[dashedgroup, fit=(mpt) (bk)] (grp_chain) {};

\node[rightbox, right=10mm of grp_phash] (lab_phash) {pHash\\Computation};
\node[rightbox, right=10mm of grp_chain] (lab_chain) {Query\\Image};

\node[decision, below=11mm of bk, minimum width=30mm, minimum height=14mm] (dec) {$d_{\min}$};

\draw[arrow] (bk) --
  node[midway, right=3mm, xshift=8mm, align=left, fill=white, inner sep=1pt]
  {$h_q$: pHash of uploaded image\\
   $h_i$: pHash from registry\\
   $d = d_H(h_q,\; h_i)$}
  (dec);

\coordinate (base) at ($(dec.south)+(0,-14mm)$);

\coordinate (leftdrop)  at ($(dec.west)+(-12mm,0)$);
\coordinate (rightdrop) at ($(dec.east)+(12mm,0)$);

\node[sidebox, anchor=north] (exact) at ($(base)+(-32mm,0)$)
{\textbf{Exact Match}\\(Flagged as registered)};

\node[sidebox, anchor=north] (potential) at (base)
{\textbf{Potential Match}\\Assign Score $S(A,B)$};

\node[sidebox, anchor=north] (nonmatch) at ($(base)+(32mm,0)$)
{\textbf{Non-Match}\\(Flagged as Not-registered)};

\draw[arrow] (dec.south) --
  node[right=1mm, xshift=6mm] {$0<d\le\tau$}
  (potential.north);

\draw[arrow] (dec.west) --
  node[above, xshift=6mm] {$d=0$}
  (leftdrop) |- (exact.north);

\draw[arrow] (dec.east) --
  node[above, xshift=6mm] {$d>\tau$}
  (rightdrop) |- (nonmatch.north);

\end{tikzpicture}%
}

\caption{Flowchart for Verification}
\label{fig:verification_flow}
\end{figure}

\subsection{Performance Analysis}
By fixing the prefix size to 4, we obtain a 16-bit prefix space, resulting in a total of $2^{16} = 65,536$ distinct buckets. During a typical query, only 137 of these buckets are searched, given the bit flip tolerance is 2, significantly reducing the search scope. Assuming a dataset containing 1 million perceptual hashes, this yields an average of approximately 15 hashes per bucket. 
One of the key advantages of BK-trees is their ability to always return the closest match to a query hash, regardless of the radius provided. Even when the search threshold $\tau$ (Hamming distance) is high such as 5 or 10, the BK-tree algorithm continues until the best match is found within the desired distance is located. This property ensures that the most similar image hash is efficiently retrieved.
This multi-stage filtering process demonstrates substantial efficiency gains over exhaustive search methods by combining prefix bucketing with approximate nearest-neighbor search using the BK-tree structure.
We analyse 4 scenarios with increasing number of pHashes to query:

\begin{table}[htbp]
\caption{Theoretical extrapolation of candidate search space (analytical; not experimentally validated).}
\centering
\begin{tabular}{c c c c}
\hline
\textbf{pHashes} & \textbf{Avg per bucket} & \textbf{Buckets searched} & \textbf{Candidates checked} \\
\hline
1M   & $\sim$15     & 137 & $\sim$2055  \\
1B   & $\sim$15259  & 137 & $\sim$2.09M \\
1T   & $\sim$15.26M & 137 & $\sim$2.09B \\
10T  & $\sim$156.2M & 137 & $\sim$20.9B \\
\hline
\end{tabular}
\label{tab:phash_comparison}
\end{table}

These values are theoretical extrapolations provided to characterize asymptotic scaling behavior of the indexing strategy; they do not imply that registries of this magnitude are currently practical or evaluated in this work.

\paragraph{Time Complexity}
The end-to-end verification process consists of several stages with distinct computational costs:

\textbf{Hash computation.} Computing the perceptual hash of a query image incurs constant time with respect to the registry size and depends only on image resolution.

\textbf{Prefix enumeration and lookup.} Prefix extraction and enumeration of 1- and 2-bit variations involve a constant number of operations, yielding at most 137 candidate prefixes. Lookup of each prefix key in the logical trie structure incurs time proportional to the prefix length, which is constant for a fixed configuration.

\textbf{BK-tree search.} Let $m$ denote the number of perceptual hashes within a candidate bucket. In the worst case, BK-tree search may degrade to $O(m)$ when the distance threshold is large or the hash distribution is highly clustered. In practice, the average-case complexity is substantially lower and depends on the distribution of hashes and the selected Hamming distance threshold, as the triangle inequality prunes large portions of the search space.

\textbf{Verification overhead.} Verification of the MPT cryptographic commitments introduces an additional logarithmic component associated with traversing the authenticated trie structure. This overhead depends on the trie depth and remains independent of the total number of registered images for a fixed prefix configuration. If the MPT root is subsequently anchored to a blockchain, blockchain-specific verification and transaction overhead would constitute an additional deployment-dependent cost that is outside the scope of this evaluation.

\subsection{Threat Model}
The proposed framework assumes an adversarial environment typical of large-scale online content platforms. Adversaries are assumed to have full access to published images and may apply arbitrary post-generation transformations, including resizing, compression, cropping, color manipulation, noise injection, and manual editing, prior to re-upload. Adversaries may also adaptively attempt to evade detection by exploiting known limitations of perceptual hashing, including targeted collision or near-collision attacks.

The framework does not attempt to identify AI-generated images produced by non-cooperative generators that do not participate in the registry, nor does it aim to resist fully adaptive adversaries capable of generating novel images specifically optimized to collide with registered perceptual hashes. Within these bounds, the system is designed to provide reliable and auditable provenance verification for registered AI-generated images under benign and moderately adversarial transformations.

\subsection{Possible Limitations}
The proposed framework operates under the assumption that participating AI image generation platforms register perceptual hashes of generated images at creation time. This assumption aligns with emerging provenance initiatives and platform-level policy enforcement, but it does not hold universally across all generative systems. Non-cooperative or malicious generators that intentionally avoid registration represent an inherent limitation of registry-based provenance approaches. Images generated by such systems cannot be verified using the proposed mechanism unless they are subsequently registered through external means. As a result, the framework should be viewed as complementary to, rather than a replacement for, forensic analysis and AI-based detection techniques that aim to classify previously unseen synthetic content.

Accordingly, the system is designed to verify whether an uploaded image matches previously registered AI-generated content or its near-duplicate variants. It does not aim to detect AI-generated images produced by non-cooperative generators that do not participate in the registry.

Failure modes of the proposed approach include: (i) false negatives for AI-generated images that are not registered, (ii) false positives arising from perceptual similarity between unrelated images, and (iii) evasion through extreme transformations or adversarial manipulation of image content such as images being flipped, rotated or heavily edited. These limitations are inherent to similarity-based and registry-driven systems and motivate the use of layered verification pipelines in practice.

Managing and querying trillions of perceptual hashes necessitates the use of distributed storage architectures, along with efficient pruning and index maintenance strategies to ensure both performance and reliability at scale.

Additionally, like other perceptual hashing schemes, pHash can be vulnerable to adversarially crafted collisions, where an attacker intentionally edits content to manipulate the resulting hash, potentially degrading verification reliability under targeted evasion.

\section{Evaluations and Results}
\subsection{Implementation}
We implemented the proposed MPT + BK-tree–based provenance verification framework in Python to evaluate both detection performance and scalability characteristics. All perceptual hashes (pHashes) were computed using a standard 64-bit pHash pipeline. The registry was populated exclusively with intact AI-generated images, while edited AI images and real images were used only at query time along with AI images, reflecting the registry-based verification assumptions of the system.

Two categories of experiments were conducted. 

First, the parameter sweep experiment. We evaluated verification accuracy using a binary classification setting under varying operational parameters. Confusion matrices were computed for multiple Hamming distance thresholds ($\tau$) and for two prefix search tolerances (2-bit and 4-bit flips), where the positive class corresponds to images present in the registry (intact and edited AI images) and the negative class corresponds to images not in the registry (real images). This experiment quantifies the trade-off between recall and false positive rate as matching tolerance is increased.

We used the AI-Generated Images vs. Real Images dataset \cite{b16}, curated by Tristan Zhang and publicly available on Kaggle. The dataset contains a balanced set of 30,000 intact AI-generated images and 30,000 real images. The intact AI-generated images are evenly distributed across three generative models: Stable Diffusion (10,000 images), MidJourney (10,000 images), and DALL·E (10,000 images). The real image set consists of 22,500 photographs sourced from Pexels and Unsplash, along with 7,500 artworks obtained from WikiArt.

To construct the Edited AI Images query set, a randomly selected subset of 6,000 intact AI-generated images was subjected to a range of image transformations, including blur, sharpening, edge enhancement, brightness and contrast adjustment, color modification, text overlay, and noise injection. These transformations were applied in varying combinations to generate a total of 30,000 edited AI images. The MPT+BK-tree registry was populated exclusively with the 30,000 intact AI-generated images, which served as the registered reference set for all subsequent verification and retrieval experiments.

Second, the performance and scalability experiment. We evaluated the scalability and query performance of the proposed indexing structure. To isolate search behavior from image content, we generated synthetic datasets of 100 K, 500 K, and 1 M random perceptual hashes. These hashes were inserted into three alternative data structures: (i) a flat array with linear lookup, (ii) a standalone BK-tree, and (iii) the proposed MPT-bucketed BK-tree structure. For each dataset size, a fixed set of 50 random query hashes was issued, and average query latency was measured across all three approaches. This experiment highlights the efficiency gains achieved by prefix-based bucketing combined with approximate nearest-neighbor search, particularly at larger scales.

Together, these experiments validate both the correctness of the registry-based verification mechanism and the scalability advantages of the proposed hybrid indexing architecture.

The implementation was conducted on hardware with the following specifications:\\
CPU: Intel(R) Core(TM) i5-9600K CPU @ 3.70GHz \\
RAM: 16GB DDR4 \\
GPU: NVIDIA GeForce RTX 2060 \\
Operating System: Windows 11 Pro 64-bit (10.0, Build 26100)

\subsection{Evaluation}

\noindent\textbf{Experiment 1: Registry Verification and Classification.}
For each query image $q$, the system performs provenance verification against the perceptual hash registry. First, a perceptual hash is computed as $h_q = \text{pHash}(q)$. The query hash is then used to search the registry by identifying candidate buckets through the prefix-based Merkle Patricia Trie (MPT), followed by similarity search within the corresponding BK-trees. For all candidate perceptual hashes $h_i$ examined during this process, the minimum Hamming distance is computed as
\[
d_{\min}(q) = \min_i d_H(h_q, h_i).
\]

If a match is found, the associated image-level metadata stored alongside the perceptual hash within the BK-tree may optionally be retrieved. A threshold-based decision rule is applied to determine registry membership: if $d_{\min}(q) \leq \tau$, the query image is classified as \textit{ found-in-registry}; otherwise, if $d_{\min}(q) > \tau$, it is classified as \textit{not-in-registry}. This experiment evaluates the effect of different Hamming distance thresholds and prefix bit-flip tolerances (2-bit and 4-bit) on verification accuracy.

Ground truth labels are defined based on registry membership rather than semantic image origin. Images whose original perceptual hashes were registered in the system are treated as the positive class ($y = 1$), while images not present in the registry are treated as the negative class ($y = 0$). Accordingly, intact and edited AI-generated query images are labeled as positive, while real image queries are labeled as negative.

The results for discontiguous 4-bit prefix are provided in Table II and continuous 4-bit prefix is provided in Table III.

\vspace{0.5em}
\noindent\textbf{Experiment 2: Scalability and Search Performance.}
To evaluate the scalability and query efficiency of the proposed indexing architecture, we conducted a controlled experiment using synthetically generated perceptual hashes. A total of one million (1M) random 64-bit hash values were generated to approximate large-scale registry conditions. From this corpus, three progressively larger subsets of size 100k, 500k, and 1M were constructed. Each subset was indexed using three alternative data structures: (i) a flat array with linear scan, (ii) a standalone BK-tree, and (iii) the proposed hybrid structure combining prefix-based bucketing with BK-trees. Following index construction, 50 query hashes were evaluated, consisting of 25 Registered hashes sampled from the corresponding registry and 25 Not-Registered hashes that were absent from the registry. The same balanced query composition was used across all three registry sizes to ensure a consistent basis for latency comparison. For every query, the minimum Hamming distance match was retrieved, and end-to-end query latency was measured. This experimental design isolates the impact of index structure on similarity search performance while controlling for hash distribution and query complexity. The results demonstrate that while linear scan and standalone BK-tree performance degrades significantly as the registry size grows, the proposed structure maintains stable and sublinear query times by aggressively pruning the search space through prefix bucketing, confirming its suitability for large-scale, low-latency similarity search provenance verification deployments.
The results are provided in Table IV.

\subsection{Results}
\noindent\textbf{Experiment 1: Registry Verification and Classification:} Table II reports confusion matrices and derived metrics for the discontinuous-prefix configuration using both 2-bit and 4-bit prefix expansion. The results demonstrate that registry-based verification achieves high recall on AI-derived images, including edited variants, while maintaining low false positive rates on real images at conservative thresholds. An operating point around $\tau$ = 6 consistently recovers approximately 98–99\% of AI-derived content with FPR below 0.2\% across both prefix configurations. Increasing $\tau$ beyond this range yields diminishing recall gains and rapidly degrades precision due to false positives, underscoring the importance of policy-driven threshold selection.

Compared to discontinuous prefixing, continuous prefixing in Table III at $\tau$=6 gives slightly higher recall but also slightly higher FPR, and for $\tau$ $>=$ 10 the FPR grows sharply

Across both 2-bit and 4-bit prefix expansion, discontinuous prefixing achieves comparable recall to continuous prefixing at conservative thresholds while consistently yielding lower false positive rates. Given the sensitivity of large-scale content moderation systems to false positives, discontinuous prefixing represents a potentially preferable configuration for deployment scenarios with strict false-positive requirements.

Across both continuous and discontinuous prefixing, increasing prefix expansion from 2-bit to 4-bit yields marginal recall gains at the cost of a consistently higher false positive rate, indicating that 2-bit expansion is the more preferable for deployment scenarios, operating choice under conservative moderation constraints.

\noindent\textbf{Experiment 2: Scalability and Search Performance:}
Table IV reports average and 95th-percentile query latency for flat array, standalone BK-tree, and the proposed Patricia Trie–based indexing structure across registry sizes up to one million hashes. Flat array search exhibits linear growth in both average and tail latency as registry size increases. Standalone BK-tree indexing degrades average performance relative to linear scan and suffers from severe tail latency. In contrast, the proposed Patricia Trie + BK-tree structure maintains consistently low average and tail latency across all registry sizes, remaining sub-millisecond at one million entries. This behavior arises from prefix-based partitioning, which bounds the candidate search space prior to metric-tree traversal.

\noindent\textbf{Summary of Experimental Findings:}
This work demonstrates that effective and scalable provenance verification of AI-generated images requires both algorithmic robustness and careful systems design. From an accuracy and robustness perspective, the registry-based verification framework reliably identifies AI-derived images including edited variants while maintaining low false positive rates on non-registered content when operated at conservative thresholds ($\tau$ = 6). The evaluation shows that discontinuous prefixing provides comparable recall to continuous prefixing while offering greater stability against false positives, and that 2-bit prefix expansion represents a more deployment-friendly default than 4-bit expansion due to its lower false positive rate under equivalent operating conditions. Together, these results emphasize that threshold selection and prefix design are policy-sensitive decisions that directly affect moderation outcomes.

From a scalability and performance perspective, the results demonstrate that similarity search performance is dominated not by average latency but by tail latency. Flat arrays are fundamentally constrained by linear scan behavior, and standalone BK-trees, while reducing average search cost, exhibit severe worst-case latency at scale. In contrast, prefix-partitioned metric indexing fundamentally alters the performance profile. By combining Patricia Trie–based bucketing with BK-tree similarity search, the proposed architecture bounds the candidate search space and achieves predictable, low-latency verification even at million-scale registries, with sub-millisecond p95 query times. Although this design incurs modest additional insertion overhead, that cost is amortized in read-heavy verification pipelines typical of large content platforms.

Taken together, these findings establish that scalable AI content provenance cannot be achieved through similarity search alone, nor through accuracy optimization in isolation. Instead, it requires joint control of false positives, robustness to benign transformations, and tail latency at scale. By integrating perceptual hashing, prefix-based partitioning, and metric-tree search within a blockchain-anchored registry, the proposed system provides a practical, transparent, and deployable foundation for large-scale AI image provenance verification. The framework complements existing watermarking and learning-based detection methods and is suitable for latency-sensitive moderation workflows and platform-level adoption.

\begin{table*}[t]
\caption{Confusion Matrices and Derived Metrics for discontiguous 2 \& 4 Bit prefix }
\label{tab:confusion_metrics}
\centering

\textbf{(a) 2-bit Prefix Flips}\\[0.5em]
\begin{tabular}{c c c c c c c c}
\hline
$\tau$ & TP & FN & FP & TN & Recall (TPR) & Precision & FPR \\
\hline
2  & 54187 & 5813 & 8     & 29992 & 0.90312 & 0.99985 & 0.00027 \\
6  & 58855 & 1145 & 45    & 29955 & 0.98092 & 0.99924 & 0.00150 \\
10 & 59185 & 815  & 410   & 29590 & 0.98642 & 0.99312 & 0.01367 \\
15 & 59367 & 633  & 6657  & 23343 & 0.98945 & 0.89920 & 0.22190 \\
20 & 59999 & 1    & 29921 & 79    & 0.99998 & 0.66725 & 0.99737 \\
\hline
\end{tabular}

\vspace{1em}

\textbf{(b) 4-bit Prefix Flips}\\[0.5em]
\begin{tabular}{c c c c c c c c}
\hline
$\tau$ & TP & FN & FP & TN & Recall (TPR) & Precision & FPR \\
\hline
2  & 54187 & 5813 & 8     & 29992 & 0.90312 & 0.99985 & 0.00027 \\
6  & 59392 & 608  & 58    & 29942 & 0.98987 & 0.99902 & 0.00193 \\
10 & 59891 & 109  & 656   & 29344 & 0.99818 & 0.98916 & 0.02187 \\
15 & 59971 & 29   & 14213 & 15787 & 0.99952 & 0.80839 & 0.47377 \\
20 & 60000 & 0    & 30000 & 0     & 1.00000 & 0.66667 & 1.00000 \\
\hline
\end{tabular}

\end{table*}

\begin{table*}[t]
\caption{Confusion Matrices and Derived Metrics for Continuous 2 \& 4 bit Prefix}
\label{tab:confusion_metrics_continuous}
\centering

\textbf{(a) 2-bit Prefix Flips}\\[0.5em]
\begin{tabular}{c c c c c c c c}
\hline
$\tau$ & TP & FN & FP & TN & Recall (TPR) & Precision & FPR \\
\hline
2  & 54187 & 5813 & 8     & 29992 & 0.903116 & 0.999852 & 0.000267 \\
6  & 59161 & 839  & 55    & 29945 & 0.986017 & 0.999071 & 0.001833 \\
10 & 59554 & 446  & 449   & 29551 & 0.992567 & 0.992517 & 0.014967 \\
15 & 59696 & 304  & 8283  & 21717 & 0.994933 & 0.878150 & 0.276100 \\
20 & 60000 & 0    & 29976 & 24    & 1.000000 & 0.666844 & 0.999200 \\
\hline
\end{tabular}

\vspace{1em}

\textbf{(b) 4-bit Prefix Flips}\\[0.5em]
\begin{tabular}{c c c c c c c c}
\hline
$\tau$ & TP & FN & FP & TN & Recall (TPR) & Precision & FPR \\
\hline
2  & 54187 & 5813 & 8     & 29992 & 0.903116 & 0.999852 & 0.000267 \\
6  & 59393 & 607  & 62    & 29938 & 0.989883 & 0.998957 & 0.002067 \\
10 & 59918 & 82   & 690   & 29310 & 0.998633 & 0.988614 & 0.023000 \\
15 & 59987 & 13   & 14944 & 15056 & 0.999783 & 0.800556 & 0.498133 \\
20 & 60000 & 0    & 30000 & 0     & 1.000000 & 0.666667 & 1.000000 \\
\hline
\end{tabular}

\end{table*}

\begin{table*}[t]

\caption{Query Latency Comparison Across Index Structures at Different Registry Sizes}

\label{tab:latency_comparison}

\centering

\begin{tabular}{c l c c c}

\hline

\textbf{Registry Size} & \textbf{Index Structure} & \textbf{Queries} & \textbf{Avg Query Time (ms)} & \textbf{p95 Query Time (ms)} \\

\hline

\multirow{3}{*}{100K}
& Flat Array              & 50 & 7.52   & 10.71 \\
& BK-tree                 & 50 & 132.11 & 225.29 \\
& Patricia Trie + BK-tree & 50 & 1.05   & 2.38 \\

\hline

\multirow{3}{*}{500K}
& Flat Array              & 50 & 54.89  & 87.30 \\
& BK-tree                 & 50 & 645.75 & 1083.74 \\
& Patricia Trie + BK-tree & 50 & 3.70   & 8.03 \\

\hline

\multirow{3}{*}{1M}
& Flat Array              & 50 & 68.04   & 99.73 \\
& BK-tree                 & 50 & 1169.28 & 2411.78 \\
& Patricia Trie + BK-tree & 50 & 6.68    & 14.30 \\

\hline

\end{tabular}

\end{table*}

\section{Ethical, Legal, and Policy Considerations}
Although perceptual hashes (pHashes) are not raw images, they still constitute derived data that could be interpreted as biometric or identifying information under certain jurisdictions. Regulations such as the EU General Data Protection Regulation (GDPR) and the California Consumer Privacy Act (CCPA) may require explicit consent for the storage of such identifiers, even if they are not reversible to the original image. To ensure compliance, platforms integrating the system should implement anonymization safeguards.

The decentralized nature of the proposed registry raises questions about oversight, moderation, and updates. Without proper governance, malicious actors could flood the ledger with false entries or misleading hashes. A robust governance framework potentially based on a multi-stakeholder consortium should define entry validation procedures, dispute resolution mechanisms, and audit processes. Smart contracts could enforce these policies transparently, ensuring that no single entity can manipulate the registry.

For maximum interoperability and adoption, the proposed framework should align with ongoing industry initiatives such as the Coalition for Content Provenance and Authenticity (C2PA), W3C provenance standards, and UNESCO’s Recommendation on the Ethics of Artificial Intelligence. By adhering to these emerging standards, the system can complement existing watermarking and metadata approaches, enhancing trust while avoiding fragmentation across platforms and jurisdictions.

\section{Future Work}
This research establishes a strong foundation for AI image provenance but opens several promising avenues for extension:

\subsection{Infrastructure-Level Interoperability with Emerging Blockchain-based Trust Infrastructures}
Although the proposed framework verifies only AI-generated image provenance and does not perform human identity validation, it is architecturally compatible with emerging blockchain-based trust infrastructures such as World \cite{b17}. Both systems rely on registry-based commitments anchored on-chain and verified off-chain by consuming applications. Future work could explore infrastructure-level interoperability, including shared governance models, standardized registry interfaces, or common cryptographic primitives, while maintaining strict separation between content provenance and identity verification domains.

\subsection{Multi-Modal Deepfake Provenance}
Future iterations could expand beyond still images to detect manipulated videos and AI-synthesized audio. Similar perceptual hashing principles could be adapted for temporal and spectral domains, enabling cross-media verification of deepfake content.

\subsection{Hybrid Hashing Approaches}
Integrating pHash with deep feature embeddings derived from convolutional neural networks or vision transformers \cite{b18} may improve robustness against heavy transformations and adversarial attacks. A hybrid approach could yield higher accuracy while retaining the efficiency of the proposed indexing structure.

\subsection{Federated Blockchain Networks}
To address scalability and jurisdictional fragmentation, the verification infrastructure could be implemented as a federated network of interoperable blockchains \cite{b19}. Each participating network could maintain local autonomy while sharing verified hash entries through cross-chain bridges, enabling global coverage without a single point of failure.

\section*{Conclusion}
This paper presented a registry-based framework for verifiable provenance identification of AI-generated images, designed to operate at the point of large-scale online distribution. Rather than attempting universal synthetic image detection, the proposed system verifies whether an uploaded image matches previously registered AI-generated content (or near-duplicate variants) using perceptual hashing and a similarity search pipeline. The design combines prefix-based bucketing in a Merkle Patricia Trie with BK-tree search within buckets to enable efficient approximate matching, while minimizing on-chain storage to cryptographic commitments and maintaining full hashes and metadata off-chain for scalability.

Experimental results demonstrate that the proposed verification mechanism can recover a high fraction of registered AI-derived content, including edited variants, while keeping false positives low at conservative thresholds. Across both continuous and discontinuous prefixing configurations, an operating point around $\tau = 6$ provides a favorable trade-off (approximately 98--99\% recall with sub-0.2\% false positive rate), whereas increasing $\tau$ yields diminishing recall gains but sharply degrades precision due to rapidly increasing false positives. The evaluation further shows that discontinuous prefixing achieves comparable recall to continuous prefixing while consistently yielding lower false positive rates, and that 2-bit prefix expansion is more deployment-friendly than 4-bit expansion under typical moderation constraints.

A separate scalability experiment using synthetic registries up to one million hashes shows that index structure strongly influences both average and tail latency. Linear scan exhibits expected growth with registry size, and standalone BK-trees suffer from severe tail latency at scale. In contrast, the proposed Patricia Trie + BK-tree structure maintains consistently low average and p95 query latency, demonstrating the benefit of bounding the candidate search space prior to metric-tree traversal. Overall, the results indicate that practical AI image provenance at platform scale requires joint control of robustness to benign transformations, false positive rates, and tail latency. 

By providing a transparent and tamper-evident registry mechanism that complements watermarking and learning-based detectors, the proposed framework offers a scalable foundation for platform-level adoption of AI image provenance verification.



\begin{thebibliography}{00}
\bibitem{b1} Luo H, Li L, Li J. Digital Watermarking Technology for AI-Generated Images: A Survey. Mathematics. 2025; 13(4):651. https://doi.org/10.3390/math13040651

\bibitem{b2} Lokner Lađević, A., Kramberger, T., Kramberger, R., \& Vlahek, D. (2024). Detection of AI-generated synthetic images with a lightweight CNN. AI, 5(3), 1575-1593.

\bibitem{b3} Zhao, X., Zhang, K., Su, Z., Vasan, S., Grishchenko, I., Kruegel, C., ... \& Li, L. (2024). Invisible image watermarks are provably removable using generative ai. Advances in neural information processing systems, 37, 8643-8672.

\bibitem{b4} Li, Z., Yan, J., He, Z., Zeng, K., Jiang, W., Xiong, L., \& Fu, Z. (2025). Is Artificial Intelligence Generated Image Detection a Solved Problem?. arXiv preprint arXiv:2505.12335.

\bibitem{b5} de Ocáriz Borde, H. S. (2022). An overview of trees in blockchain technology: merkle trees and merkle patricia tries. University of Cambridge: Cambridge, UK.

\bibitem{b6} Burkhard, W. A., \& Keller, R. M. (1973). Some approaches to best-match file searching. Communications of the ACM, 16(4), 230-236.

\bibitem{b7} Farid, H. (2021). An overview of perceptual hashing. Journal of Online Trust and Safety, 1(1).

\bibitem{b8} Ahmed, N., Natarajan, T., \& Rao, K. R. (2006). Discrete cosine transform. IEEE transactions on Computers, 100(1), 90-93.

\bibitem{b9} Heidari, A., Navimipour, N. J., Dag, H., Talebi, S., \& Unal, M. (2024). A novel blockchain-based deepfake detection method using federated and deep learning models. Cognitive Computation, 16(3), 1073-1091.

\bibitem{b10} Khalaf, L. I., Jumaili, M. L. F., Almashhadany, M. T. M., Aljanabi, M. S., Hasan, T. S., \& Algburi, S. (2024, May). Image forgery detection using convolutional neural networks and blockchain technology. In Proceedings of the Cognitive Models and Artificial Intelligence Conference (pp. 316-321).

\bibitem{b11} Mastoi, Q. U. A., Memon, M. F., Jan, S., Jamil, A., Faique, M., Ali, Z., ... \& Syed, T. A. (2025). Enhancing Deepfake Content Detection Through Blockchain Technology. International Journal of Advanced Computer Science \& Applications, 16(6).

\bibitem{b12} Farid, H. (2017). Reining in online abuses. Technology and Innovation, 19(3), 249-255.

\bibitem{b13} Rosenthol, L. (2022, October). C2PA: the world’s first industry standard for content provenance (Conference Presentation). In Applications of Digital Image Processing XLV (Vol. 12226, p. 122260P). SPIE.

\bibitem{b14} Gowal, S., Bunel, R., Stimberg, F., Stutz, D., Ortiz-Jimenez, G., Kouridi, C., ... \& Kohli, P. (2025). SynthID-Image: Image watermarking at internet scale. arXiv preprint arXiv:2510.09263.

\bibitem{b15} Lashkari, B., \& Musilek, P. (2021). A comprehensive review of blockchain consensus mechanisms. IEEE access, 9, 43620-43652.

\bibitem{b16} https://www.kaggle.com/datasets/tristanzhang32/ai-generated-images-vs-real-images

\bibitem{b17} Gent, E. (2023). A Cryptocurrency for the Masses or a Universal ID?: Worldcoin Aims to Scan all the World's Eyeballs. IEEE Spectrum, 60(1), 42-57.

\bibitem{b18} Park, N., \& Kim, S. (2022). How do vision transformers work?. arXiv preprint arXiv:2202.06709.

\bibitem{b19} Nguyen, C. T., Hoang, D. T., Nguyen, D. N., Xiao, Y., Pham, H. A., Dutkiewicz, E., \& Tuong, N. H. (2023). Fedchain: Secure proof-of-stake-based framework for federated-blockchain systems. IEEE Transactions on Services Computing, 16(4), 2642-2656.

\end{thebibliography}
\end{document}